\documentclass[utf8]{FrontiersinHarvard} % for articles in journals using the Harvard Referencing Style (Author-Date), for Frontiers Reference Styles by Journal: https://zendesk.frontiersin.org/hc/en-us/articles/360017860337-Frontiers-Reference-Styles-by-Journal
\usepackage[colorlinks=true, linkcolor=blue, citecolor=blue, urlcolor=blue]{hyperref}
\usepackage{url,hyperref,lineno,microtype,subcaption}
\usepackage[onehalfspacing]{setspace}

\def\keyFont{\fontsize{8}{11}\helveticabold }
\def\firstAuthorLast{Giri {et~al.} 2024} %use et al only if is more than 1 author {et~al.}
\def\Authors{Gourab Giri\,$^{1,2,*}$, Christian Fendt\,$^{3}$, Kshitij Thorat\,$^{1}$, Gianluigi Bodo\,$^{4}$, Paola Rossi\,$^{4}$}
\def\Address{$^{1}$Department of Physics, University of Pretoria, Private Bag X20, Hatfield 0028, South Africa \\
$^{2}$South African Radio Astronomy Observatory, 2 Fir St, Black River Park, Observatory 7925, South Africa \\
$^{3}$Max Planck Institute for Astronomy, Königstuhl 17, D-69117 Heidelberg, Germany \\
$^{4}$INAF/Osservatorio Astrofisico di Torino, Strada Osservatorio 20, 10025 Pino Torinese, Italy}
\def\corrAuthor{Gourab Giri; Christian Fendt}

\def\corrEmail{gourab.giri@up.ac.za; fendt@mpia.de}

\begin{document}
\onecolumn
\firstpage{1}

\title[XRGs: Jet Evolution and Global Dynamics]{X-Shaped Radio Galaxies: Probing Jet Evolution, Ambient Medium Dynamics, and Their Intricate Interconnection} 

\author[\firstAuthorLast ]{\Authors} %This field will be automatically populated
\address{} %This field will be automatically populated
\correspondance{} %This field will be automatically populated

\extraAuth{}% If there are more than 1 corresponding author, comment this line and uncomment the next one.
%\extraAuth{corresponding Author2 \\ Laboratory X2, Institute X2, Department X2, Organization X2, Street X2, City X2 , State XX2 (only USA, Canada and Australia), Zip Code2, X2 Country X2, email2@uni2.edu}

\maketitle

\begin{abstract}
This review explores the field of X-shaped radio galaxies (XRGs), a distinctive subset of winged radio sources that are identified by two pairs of jetted lobes which aligned by a significant angle, resulting in
an inversion-symmetric structure. 
These lobes, encompassing active (primary) and passive (secondary) phases, exhibit a diverse range of properties across 
the multiple frequency bands, posing challenges in discerning their formation mechanism. 
The proposed mechanisms can broadly be categorized into those related either to a triaxial ambient medium, into which the jet propagates, or to a complex, central AGN mechanism, where the jet is generated. 
The observed characteristics of XRGs as discovered in the most substantial sample to date, 
challenge the idea that there is universal process at work that produces the individual sources of XRGs.
Instead, the observational and numerical results rather imply the absence of an universal model and infer that distinct mechanisms may 
be at play for the specific sources.
By scrutinizing salient and confounding properties, this review intends to propose the potential direction for future research 
to constrain and constrict individual models applicable to XRGs.

\section{}

\tiny
 \keyFont{ \section{Keywords:} galaxies: active, (galaxies:) quasars: supermassive black holes, galaxies: jets, ISM: jets and outflows, radio continuum: galaxies, X-rays: galaxies: clusters, galaxies: clusters: intracluster medium, (magnetohydrodynamics) MHD} %All article types: you may provide up to 8 keywords; at least 5 are mandatory.
\end{abstract}

\section{Introduction} 

The examination of emission spectrum observed from a galaxy allows for the categorization of galaxies into two primary types: normal and active galaxies. The emission from the core of a normal galaxy primarily arises from stars characterized by a black-body spectrum and is comparable to the emission from the rest of the galaxy. However, in active galaxies, the emission from the central region (Active Galactic Nuclei; AGN) is much higher, $\sim$ 100 - 1000 times greater than the emission from other regions of the galaxy, and produces a distinctive double-hump non-thermal emission spectrum \citep{Elvis1994}. The active galaxies are identified with Eddington ratios exceeding the limit of \(L_{\text{AGN}}/L_{\text{EDD}} = 10^{-5}\). Here, \(L_{\text{AGN}}\) is the bolometric luminosity of the AGN, and \(L_{\text{EDD}} = 1.5 \times 10^{38} M_{\text{BH}}/M_{\odot}\ \text{erg s}^{-1}\) is the Eddington luminosity ($M_{\text{BH}}/M_{\odot}$ represents the mass of the central massive object, identified as black hole, in solar mass) \citep{Urry1995}. Observations of the central regions of nearby galaxies, including ellipticals, lenticulars, and spiral bulges, reveal the presence of a supermassive black hole (SMBH; with mass $M_{\text{BH}} \gtrsim 10^6 M_{\odot}$) in nearly all of these sources \citep{Tremaine2002,Graham2012}. Galaxy mergers, in this regard, rejuvenate galaxies by supplying fresh gas and dust, potentially triggering AGN activity around the SMBH, which accretes surrounding matter through an accretion disk \citep{DiMatteo2005,Cotini2013,Capelo2015,Elison2019}. An actively accreting SMBH at the center of the AGN therefore emerges to be the source of such non-thermal emission \citep{Padovani2017}. 

The evolution of the central black hole and the activity of the AGN are intricately linked to the availability of material in the central regions of the galaxy. In the context of hierarchical galaxy formation models, contemporary galaxies emerge from successive mergers of smaller galaxies \citep{Toomre1972,Bullock2005,Mancillas2019}. During these (gas rich) galaxy mergers, gas is funneled toward the center on a timescale of approximately $10^8$ years, thereby initiating starbursts and AGN activity \citep{Gaskell1985,Hernquist1995,Barnes1996}. A widely accepted paradigm posits that such interactions induce the inward flow of gas from the outer regions of a galaxy to its central areas, facilitated by the loss of angular momentum triggered by tidal forces \citep{Mihos1996}. Evidence of such deep links between galaxy mergers and nuclear activity has been found in a variety of active galaxies \citep[e.g.,][]{Keel1985,Wilson1995,Springel2005,Silverman2011,Satyapal2014,Goulding2018}.

Active Galactic Nuclei are classified into two categories based on their emission characteristics in the radio band: radio-loud (RL) or radio-quiet (RQ) AGNs. This categorization is established by the ratio of radio to optical flux ($R = S_{\rm 5GHz}/S_{\rm B-Band}$) \citep{Kellermann1989}. AGNs are designated as radio-loud if $R$ $\geq 10$, while those with $R$ $< 10$ are categorized as radio-quiet. The distinction between the two populations can also be made based on radio luminosity. Radio-quiet AGNs are characterized by lower luminosities ($L_{\rm 6GHz}$ between $10^{21}$ and $10^{23.2}$ W Hz$^{-1}$), whereas radio-loud AGNs exhibit higher luminosities ($L_{\rm 6GHz}$ greater than $10^{23.2}$ W Hz$^{-1}$) \citep{Kellermann2016}.

Among radio-loud AGNs, a notable feature is the presence of (sub)relativistic jets extending from a few parsecs (pc) to megaparsec (Mpc) scales, perpendicular to the underlying accretion disk \citep{Blandford2019,Hardcastle2020}. The jets are believed to be formed through the interplay of magnetic field and rotation either of the black hole \citep{Blandford1977} or of the accretion disk \citep{Blandford1982}, where a fraction of hot and ionized matter accreting onto the supermassive black hole is expelled at high velocities. Approximately 15-20 percent of all AGNs are identified as radio-loud based on the findings by \cite{Kellermann1989}. However, the study by \cite{Padovani2011} indicate an even lower fraction of AGNs exhibiting jet activity. Detected over various scales, ranging from sub-kpc to several Mpc distances \citep{Kharb2019,Dabhade2020,Webster2021,Oei2022}, these jetted outflows may terminate within the host galaxy or may extend to the larger scales of a galaxy cluster. The seminal work by \citet{Fanaroff1974} introduced two classifications of extended jets based on their radio power: Fanaroff \& Riley (FR) class I and II. Identification of these extended radio galaxies into these categories is possible by observing the absence (lower jet power) or presence (higher jet power) of an edge-brightened feature at the jet termination point, respectively. Studies also indicate the existence of another category of compact sources featuring jetted structures on a parsec scale, generally characterized by greater symmetry and displaying mildly relativistic behavior; these sources belong to the FR 0 category, the most abundant class of radio galaxies in our local Universe, distinguished by higher core dominance \citep{Baldi2023}.

Once losing their collimation, the jets exhibit a diverse range of sub-structures, influenced by interactions with the surrounding environment or by  the internal dynamical configurations. Notable among these sub-structures is the phenomenon of jet bending \citep[e.g.,][]{Bridle1994,Krause2019,Rodman2019,Bruni2021}. In a subset of these extended radio galaxies, the jets display significant bending, deviating markedly from their initial propagation direction. This bending introduces additional complexity in determining the origin of these jettted structures, leading to the emergence of distinct and peculiar bent jetted sources.

In a small but significant subset of double-lobed radio galaxies, a distinctive deformation is evident in their lobes, giving rise to peculiar radio structures \citep{Leahy1992,Saripalli2018,Bhukta2022B}. These structures can be categorized into two primary types based on their bent jetted morphologies \citep{Ekers1982}. The first type, mirror symmetric sources, features jetted lobes bending away from the central galaxy in the same direction, forming elongated features resembling tails, known as `tailed sources'. The second type, inversion symmetric sources, exhibits lobes bending in opposite directions, resulting in the formation of `winged sources' characterized by extended structures resembling wings.

Tailed radio galaxies can manifest as `C', `U' or `V' shapes, with twin-tailed configurations frequently observed \citep[e.g.,][]{Mao2010,Muller2021,Bhukta2022B}. Tailed sources are in general classified based on the alignment angle of the twin-tailed jetted lobes, leading to distinctions such as narrow-angle tails (NAT) and wide-angle tails (WAT). NAT radio galaxies display parallel, tightly collimated tails \citep{Sebastian2017}, while WAT radio galaxies exhibit broader tails fanning out from the core \citep{Odea2023}. The consensus is that such mirror-symmetric sources form when the galaxy is in motion relative to the ambient medium or due to the wind flow of the ambient environment resulting from internal turbulence. In most instances, the ram pressure of the cluster medium and buoyancy forces contribute to the formation of these tailed structures \citep{Smolcic2007,Oneill2019,Rudnick2021,Pandge2022}.

Winged radio galaxies showcase distinctive morphologies, including `X', `S', `Z', or `W'-like structures \citep{Lal2019,Yang2019,Bera2020,Bhukta2022}, arising from the bending of jet lobes in opposite directions. This phenomenon is also observed in several microquasars \citep{Roberts2008,Marti2017}. Distinguishing between different shapes of winged sources in low-resolution observations can be challenging. However, advancements in telescope sensitivity and resolution, particularly in low-frequency radio observations, facilitate the discrimination of these morphologies \citep{Cotton2020,Bruni2021}. In these radio galaxies, one pair of lobes undergoes active evolution as an active jet carves out the lobe, often resulting in hotspot formation. These lobes are termed active or primary lobes. The other pair of lobes, typically diffuse, extended, and of lower luminosity, lack hotspots and are designated as wing or secondary lobes.

\begin{figure*}[h!]
\centering
\includegraphics[width=\columnwidth]{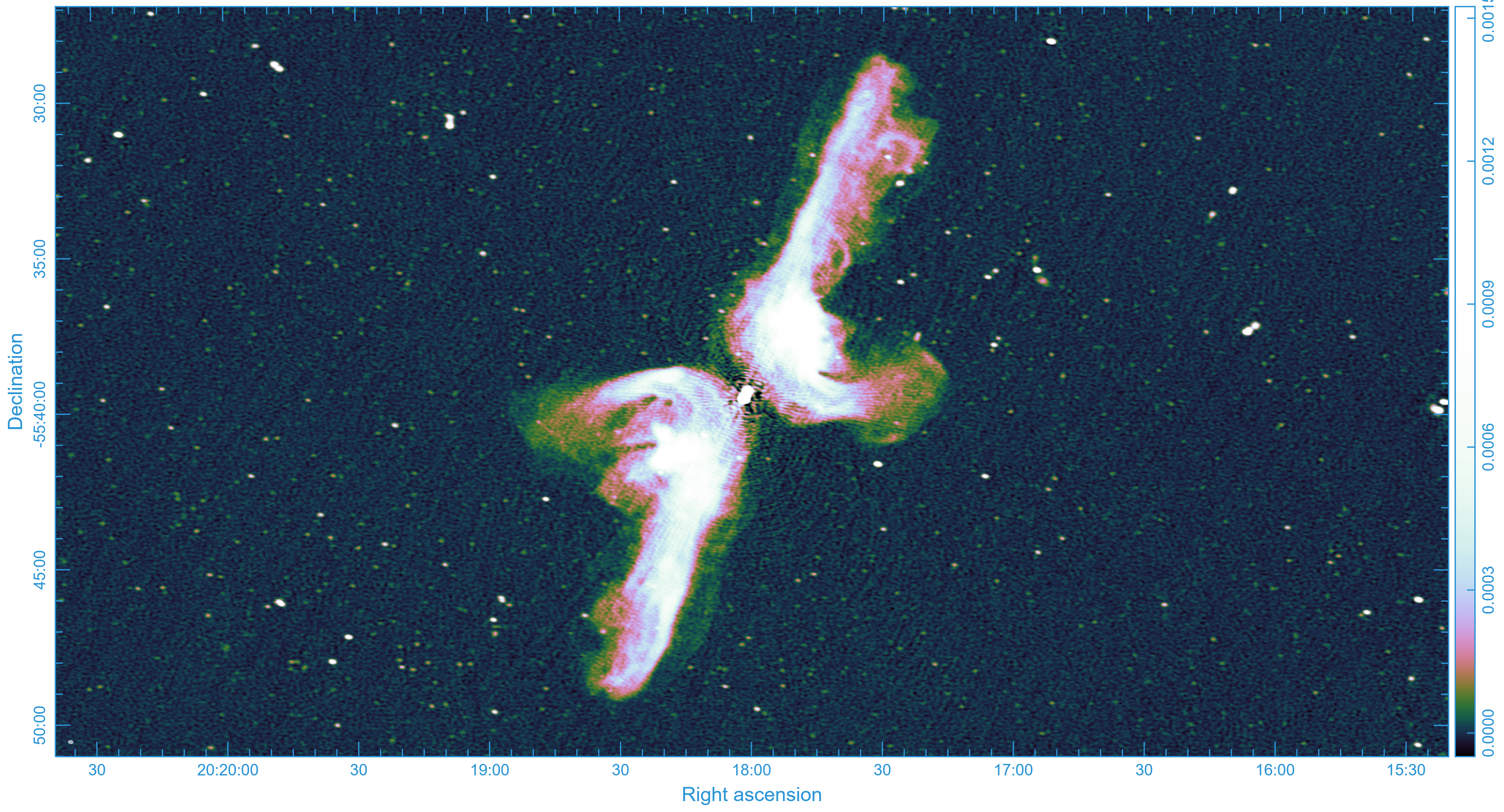}
\caption{L-band MeerKAT image of PKS 2014-55, an extended radio galaxy exhibiting a double-lobed structure reminiscent of a classic X-shaped configuration. The intricate double boomerang morphology reveals the presence of several complex structures, suggesting an enriched evolutionary mechanism associated with this source. While the giant outer radio structure, spanning nearly 1.57 Mpc, is currently evolving passively, there has been a renewed AGN activity detected, making it a source with a wealth of enriched physical processes \citep{Cotton2020}. \copyright\ Thorat et al. (in prep.).}
\label{Fig:PKS2014-55}
\end{figure*}

The formation mechanism of such sources remains a subject of ongoing debate, as recent discoveries often challenge earlier conclusions regarding the sources' origin. In this context, a comprehensive review has been conducted here focusing on the formation and long-term evolution of X-shaped radio galaxies (XRGs; Fig.~\ref{Fig:PKS2014-55}), which have proven to be efficient tools for probing various aspects of jet evolution. This includes the evolution of the central supermassive black hole, jet dynamics encompassing both active and passive phases, the dynamic configuration of the ambient medium with emphasis on the influence of magnetohydrodynamical processes in shaping the jetted structure, and the investigation of large-scale jet-ambient medium interaction. Moreover, the review explores intricate connections between X-shaped radio galaxies and their relatives, such as Z- or S-shaped winged sources. This analysis contributes to constraining and understanding the formation processes of winged galaxies in a broader context. Therefore, the discussion encompasses the collective insights gained from studying XRGs, shedding light on their multifaceted role in unraveling the complex interplay between AGN activity, jet dynamics, and their impact on the surrounding cosmic environments. 

We delve into the discussion of the current understanding of such extended radio sources, commencing with Section~\ref{Sec:Dynamical Configuration of XRGs}. This involves an exploration of the reported samples of XRGs, their morphological appearance in a broader sense, properties of their host AGNs, and the ambient environment in which they reside. Thereafter, in Section~\ref{Sec:Emission Characteristics of XRGs}, we discuss the emission properties of XRGs, indicating what the intensity, spectral, and polarization mapping of XRGs tell us. Section~\ref{Sec:XRG formation models and Challenges} incorporates a comprehensive discussion on the categorization of the XRG models, including the formulation of models, their strengths, and caveats. In Section~\ref{Sec:Potential Future Prospects}, we highlight various prospects for future studies of XRGs. We summarize this review in Section~\ref{Sec:Summary}.

\section{Dynamical Configuration of XRGs} \label{Sec:Dynamical Configuration of XRGs}
Here, we dive into a discussion on the macro-scale properties of X-shaped radio galaxies, accompanied by an examination of the reported properties of their host AGNs.

\subsection{XRG Morphology} \label{Sec:XRG Morphology}
Distinctive X-shaped radio galaxies, see Fig.~\ref{Fig:PKS2014-55} for a classic example, are identified by their unique configuration, featuring two prominent doubled-lobed jetted structures aligned at a significant angle to each other. The initial identification of such sources was documented in the work conducted by \citet{Riley1972,Hogbom1974}, followed by several seminal studies on classic X-shaped sources such as \citet{Ekers1978,Leahy1984,Leahy1992,Worrall1995,Murgia2001}. The investigation of such sources has affirmed the existence of a primary pair of lobes resembling the classical double-lobed radio structure. This structure is characterized by the presence of well-collimated, actively propagating bidirectional jets. Additionally, there are secondary lobes or wings, which are diffuse, extended, and weaker compared to the active lobes. The morphology can be further categorized into two types: (a) Inner-deviation sources, where the wings are observed to be connected with the central AGN, and (b) Outer-deviation sources, where the wings are observed to originate from the end of active lobes. The former results in a distinct X-like morphology, while the latter exhibits mostly a Z- or S-like morphology. This differentiation has been explored in subsequent studies by \citet{Roberts2015,Saripalli2018}. Nevertheless, studies has not yet conclusively established distinctions based on these topological differences. Irrespective of this, in general, the wings are observed to align at a significant angle to the active lobe, with an average alignment angle of approximately $75^{\circ}$ or greater \citep{Capetti2002,Bhukta2022}.

Recent observations employing high-resolution and sensitive telescopes have introduced additional intricacies to the existing configuration of XRGs. Notable complexities include the presence of a substantial extended tail originating from one arm of an XRG \citep{Hardcastle2019}, active jet spine sometimes generating prominent intrinsic S-like structures \citep{Bruno2019,Baghel2023}, the emergence of `W'-like global structures in some of these winged sources \citep{Proctor2011,Lal2019}, the observed expansive, diffuse wing-lobe structure \citep{Sejake2023}, and the existence of an arc-like filamentary structure enveloping the X-shape \citep{Ignesti2020}.

To ascertain the general properties of X-shaped radio galaxies, a systematic search and analysis of a representative XRG sample are crucial. In this context, \citet{Leahy1992} noted that $\sim$ 10 percent of jetted galaxies in the 3CRR catalogue exhibit X-shaped morphology. Subsequently, \citet{Lal2007} conducted mapping observations of 12 XRG sources using the Giant Metrewave Radio Telescope (GMRT) at $240$ MHz and $610$ MHz, covering almost all known XRG sources at that time. Later that year, \citet{Cheung2007} compiled a catalog of 100 XRG sources identified from the Very Large Array (VLA) Faint Images of the Radio Sky at Twenty-centimeters (FIRST) survey. Utilizing an automated morphological classification scheme on FIRST radio sources, \citet{Proctor2011} identified 155 XRG candidates, with 21 sources overlapping with the \citet{Cheung2007} sample. Furthermore, \citet{Yang2019} investigated 5128 FIRST radio sources, discovering 290 new XRGs by adopting less stringent selection criteria in cataloging XRG candidates. This included sources displaying short wings (or even a one-sided wing) and those showing only a hint of X-shaped radio structure. Among these, 25 were already part of the \citet{Proctor2011} list. Recently, \citet{Bera2020} cataloged 296 winged radio sources from the same FIRST database, imposing a lower limit of $10''$ on the largest radio size. This catalog comprises 161 XRG candidates and 135 candidate ZRGs (outer deviation sources), with 21 sources already present in the \citet{Proctor2011} catalogue. Consequently, the combined XRG candidate samples from the FIRST survey itself resulted in a total of 640 XRGs. Most recently, based on the Tata Institute of Fundamental Research (TIFR) GMRT sky survey (TGSS) catalog, \citet{Bhukta2022} reported 58 additional winged sources (40 inner-deviation sources and 18 outer-deviation sources) discovered at the lower frequency of $150$ MHz, a survey that is sensitive to detection of diffuse emission from older particles in large-sized sources.

Several studies leveraging available samples have highlighted certain properties associated with XRGs that currently appear to be generally applicable. Notably, studies such as \citet{Saripalli2009,Bera2020,Bhukta2022} demonstrate that the majority of XRGs exhibit a wing-to-lobe length ratio of less than 1 (median value of 0.9). In fact, the former study suggested that sources with larger extents generally have smaller wing lengths. Additionally, investigations by \citet{Saripalli2018,Joshi2019} indicate that approximately $70-80$ percent of XRGs have wing lengths that are either lower or comparable to the active lobes. However, it is also important to emphasize the existence of sources exhibiting wings larger than the active lobes. For instance, \citet{Bruno2019} reported an XRG with a wing-to-lobe length ratio of 2.8 \citep[see][for other such examples]{Gower1982A,Wang2003,Ignesti2020}. Consequently, it is anticipated that a successful model will have to be able to describe both groups of sources. A note of caution is warranted here, as the measurement of such lengths is intricately dependent on the sensitivity of the telescope in capturing emissions from relatively cooled particles, as well as on the projection effect \citep{Hodges-Kluck2011,Yang2019,Giri2022A}.

A limited number of XRGs exhibit passive evolution in both the wing and the active lobe, suggesting that the phase of jet activity responsible for producing the structure has ceased \citep{Saripalli2009}. However, a notable fraction of such sources demonstrate a restart of jetted activity near the center, as observed in 5 out of 8 discovered sources by \citet{Saripalli2009}. Classic examples of sources with this morphological feature include PKS 2014-55 \citep{Cotton2020} and CGCG 292-057 \citep{Misra2023}. The renewed AGN activity for PKS2014-55 can be observed near the centre in Fig.~\ref{Fig:PKS2014-55}, although not very well resolved. Interestingly, in all these cases, the propagating fresh jet pair aligns along the active jet of the earlier episode, suggesting a scenario where jet reorientation to another direction, at least for XRGs, is not common.  

\subsection{Host AGN}

Various studies have delved into the nuclear regions of such radio galaxies through multi-wavelength observations, revealing, for instance, 9-14 percent quasars, a few identified broad-line radio galaxies, and blazars \citep{Wang2003,Saripalli2018,Yang2019,Baghel2023}. 

The study conducted by \citet{Mezcua2011} further analyzed the black hole masses of AGNs in 29 XRGs and compared them with a control sample of 36 radio-loud AGNs (including 6 FR type II sources) exhibiting similar redshifts, optical, and radio luminosities. Their findings revealed a higher black hole mass in XRGs compared to the control sample ($\sim$ 1.5 times higher), with 60 percent of XRGs displaying a black hole mass, ${\rm log}(M_{\rm BH}/M_{\odot})$, greater than $8.25$. In 2012, they extended this analysis to another 12 XRGs, providing further support for their conclusions \citep{Mezcua2012}. \citet{Joshi2019} thereafter discovered, through studies of 67 XRGs \citep[including 41 from][]{Mezcua2011,Mezcua2012}, an average black hole mass, ${\rm log}(M_{\rm BH}/M_{\odot})$, of $8.81$. However, when compared to a sample of normal radio galaxies of FR-II type, they found this value to be lower than the average black hole mass for FRIIs, which is $9.07$ (in ${\rm log}(M_{\rm BH}/M_{\odot})$). A diverse range of SMBH masses of XRG host galaxies has been noted further in \citet{Liu2012} ranging from ${\rm log}(M_{\rm BH}/M_{\odot})$ of 7.05 for J1348+4411 to 9.08 for J1614+2817. 
Therefore, these studies do not conclusively suggest any discernible pattern regarding the mass characteristics of SMBHs in XRGs compared to straight bidirectional jetted sources, understanding of which could offer valuable constraints in refining formation models. A caution should also be given to the method used to estimate black hole masses \citep[e.g., see][]{Tremaine2002} which is generally applied to bulge-dominated systems, potentially leads to underestimated mass calculations in systems without a well-defined bulge, such as galaxies undergoing mergers \citep{Koziel2012}. 

Studies have also been conducted to explore the possibility of binary/dual SMBHs in the XRG host galaxies, considering separations on the parsec/kiloparsec scale. The hypothesis is that the lobes and wings may signify two separate jet episode events, possibly resulting from a jet reorientation activity. An avenue for investigating this involves examining whether the curved morphology observed in several XRGs indicates any jet precession activity. This investigation entails modeling the morphology, for example, using kinematic jet precession models \citep{Hjellming1981, Gower1982B}. Studies, exemplified by observational research from \citet{Gower1982A, Gong2011, Rubinur2017, Krause2019, Nandi2021}, along with pivotal simulation contributions by \citet{Horton2020, Giri2022B}, have explored different aspects of such cases, e.g., mass ratios, precession period and separation of the SMBH pair, shedding light on the potential parameter space for such binaries.

Identifying potential dual/binary AGN candidates through observations is a formidable challenge in general. For instance, the optical nuclear spectra often exhibit double-peaked AGN (DPAGN) emission lines, considered a potential observational signature of dual AGN \citep{Fu2011}. However, the origin of these double peaks can vary, with possible explanations including jet-medium interactions or the presence of a rotating gaseous disk \citep{Fu2012, Kharb2015, rubinur2019, Kharb2019}. In the context of XRGs, the presence of a dual-peak signature has been documented in studies by \citet{Zhang2007,Koziel2012,Rubinur2017,Nandi2021}. 
Furthermore, identification of unusually broadened emission lines in the optical spectra of AGNs may also suggest the presence of a binary black hole system formed through a recent merger \citep{Peterson1987,Gaskell1996}. Despite this, the search for such signatures in AGNs associated with XRGs has revealed that this is not a widely observed characteristic \citep{Landt2010}. Nevertheless, the presence of a few XRGs exhibiting this signature necessitates further investigation \citep{Wang2003, Cheung2007, Zhang2007}. 

\citet{Landt2010} additionally demonstrated that the nuclear regions of XRGs lack a dusty environment and have higher temperatures, providing evidence against the occurrence of any recent (gas rich) merger.
However, counterarguments challenging this assertion can be posed based on the presence of XRGs exhibiting a high-excitation spectral signature in their optical nuclear spectra. The primary diagnostic for high excitation is the [O III] $\lambda 5007$ line luminosity (with a $\lambda 5007$ line equivalent width exceeding $5\ {\rm \AA}$) \citep{Best2012}. These high-excitation radio galaxies consistently display blue colors in the color–color diagram, indicative of recent star formation triggered by a wet merger \citep{Baldi2008, Smolcic2009}. Among XRGs, a notable fraction \citep[$\sim 50$ percent;][]{Gillone2016, Joshi2019} exhibit a high-excitation state, while the remainder are categorized as low-excitation XRGs, featuring redder optical colors. Although such high-excitation radio properties are typical of FR IIs \citep{Buttiglione2010}, implying a recent gas-rich merger, further investigation is required to determine the recency of the merger. \citet{Gillone2016} demonstrated that the age of formed young stars can extend up to several Gyr, surpassing the typical age of a radio galaxy \citep{Herwood2017}. Nevertheless, a significant subset of XRGs has been identified to exhibit recent starburst activity ($< 10^6$ yr), as will be elucidated in the next section. The elevated occurrence of this excitation state in XRGs, signifying recent mergers, is notable, yet explicit merger signatures in such galaxies are scarce. Detecting merger signatures is challenging due to their low luminosity compared to the host galaxy, requiring deep imaging and specialized techniques for revealing obscured asymmetries \citep{Mancillas2019, Giri2023A}. Additionally, minor galaxy interactions have the potential to contribute significant amounts of gas to the primary galaxy without leaving a distinct signature on the host galaxy \citep{Dennett-Thorpe2002,Kaviraj2014}. Further investigations are essential to address the observed dichotomy in XRGs, where an almost equal fraction displays high and low excitation classes, offering insights that may either challenge or support proposed formation models.

Relevant X-ray and radio-VLBI observations have been undertaken in several studies to gain a deeper understanding of the cores of XRGs. The X-ray study led by \citet{Hodges-kluck2010A} reported the detection of X-ray jets near the cores of several XRG systems. Notably, their investigation revealed no evidence of misaligned additional pairs of jets in any of the XRG samples \citep[see also,][]{Miller2009}, posing a preliminary challenge to the theory put forth by \cite{Lal2007} that suggested binary or dual SMBHs at the center could be in charge of ejecting two bidirectional jets at an angle, resulting in the formation of X-morphology. Radio observations of a few sources' cores also failed to detect evidence of bidirectional jets emerging from binary systems \citep{Murgia2001,Nandi2021}, with the exception of \citet{Yang2022}. However, the latter study does not immediately clarify how such largely separated cores eventually produce the primary X-shaped morphology.

\subsection{Ambient Environment}

The host galaxies of XRGs predominantly belong to the category of elliptical galaxies, characterized by a high degree of ellipticity \citep{Capetti2002}. The gaseous distribution within these galaxies (i.e., the inter-stellar medium) often exhibits an even higher ellipticity compared to their stellar distribution \citep{Kraft2005,Hodges-kluck2010A}. This ellipsoidal ambient environment trend also extends to the intra-group and intra-cluster medium, aligning with the distribution of the host galaxy \citep{Hodges-kluck2010A,Hodges-Kluck2011}. Furthermore, in comparison with FR II host galaxies, the hosts of XRGs exhibit higher median ellipticity values \citep{Saripalli2009,Gillone2016}, thus indicating a potential correlation between XRG morphology and the ambient environment. A handful of sources, however, indicate that it is not always necessary for XRGs to have an elliptical atmosphere; for example, their hosts can exhibit a circular geometry \citep[as seen in 6 cases discovered by][]{Saripalli2009}. There are also cases in which the XRG hosts exhibit an asymmetric distribution of ambient environment, induced by a potential recent galaxy merger \citep{Heckman1982, Evans1999, Hodges-kluck2010B, Misra2023} or a possible cluster merger \citep{Hodges-kluck2012,Hardcastle2019}. 

Another notable trend in XRGs is the alignment of the wing structure almost along the minor axis of the ambient medium, implying that the wings are aware of the host medium's geometry \citep{Saripalli2009,Hodges-kluck2010A,Gillone2016}. For the active lobes, a spread of up to $50^{\circ}$ from the major axis of the host galaxy has been observed. In prominent XRGs, the active lobes are typically observed aligning along the major axis of the galaxy and wings along the minor axis \citep{Wang2003,Bruno2019}. A similar investigation conducted by \citet{Joshi2019} supports this claim but also led to the identification of six XRGs exhibiting indications of counterexamples. In these instances, the wings are reported to align along the major axis of the ambient medium \citep[see][for individual such cases]{Hodges-kluck2010B,Yang2022}. The existence of even a limited number of such counterexamples is noteworthy, urging caution in endorsing any model that aims to explain the observed general trend. 

In the analysis of the interstellar medium of XRG host galaxies using mid-infrared color measurements, \citet{Joshi2019} identified a notable presence of a young star population and/or enhanced dust masses in a substantial fraction of sources ($\sim 80$ percent). This implies that the replenishment of gas and dust to XRG hosts may have occurred through galaxy merger events. However, it remains to be determined whether these potential merger events are recent or represent older activity (older than the lobe ages). 
Other investigations, such as those by \citet{Mezcua2012}, revealed evidence of recent starburst activity in smaller fraction of XRGs (starbust that happened $10^6$ years ago), with an additional 50 percent of sources exhibiting starburst activity occurring more than $\sim 10^8$ years ago (substantially older than lobe ages). The study conducted by \citet{Gillone2016} reported even a smaller fraction of their sample ($\sim 36$ percent) displaying a young star population (with ages less than 3 Gyr). From such analysis of the interstellar medium in the host galaxies or their nuclear regions (as discussed in the previous section), the presence of dust or young stars is appearing not to be a prevalent characteristic. If present, such activity is associated with the dust inclusion or starburst events that occurred in the distant past. 

The above conclusion can further be supported by the findings of \citet{Joshi2019}, who determined that XRGs tend to inhabit low-density, large-scale environments with a median richness (i.e., number of galaxies within a projected radius of 1 Mpc and redshift bounds of $\pm 0.04 (1+z)$ centered at the source) of approximately $\sim 8.9$. In comparison, FRIIs exhibit environments with a median richness of $\sim 11.8$, and FRIs are found in environments with a median richness of $\sim 29.8$. Similar sparse environment has also been reported for XRGs by \citet{Dennett-Thorpe2002}. Given that XRGs tend to reside in poor environment does not immediately imply their host galaxy lack evolution through galaxy merger. For example, \citet{Koziel2012,Misra2023} have specifically presented an example of a winged radio galaxy residing in a low-density environment, yet exhibiting signs of a recent merger. The presence of prominent dust lanes \citep[identified as a possible merger signature, see][]{Giri2023A} has also been observed in a few XRGs, including 3C 433 \citep{Miller2009} and PKS 2014-55 \citep{Cotton2020}. Therefore, the notable fraction of XRG hosts showing signs of recent starburst or enhanced dust masses may indicate a link to recent mergers.

\section{Emission Characteristics of XRGs} \label{Sec:Emission Characteristics of XRGs}
Much investigation has been carried out to examine the X-shaped structure across different wavelengths, ranging from radio to X-ray bands, followed by seminal numerical works. In this section, we delve into the implications and discussions arising from these findings.  

\subsection{Total Intensity Continuum}

X-shaped radio galaxies place themselves near the boundary between FR I and FR II in terms of radio power \citep{Cheung2007}, tending to be more prevalent on the weaker FR II side \citep{Gillone2016,Yang2019, Saripalli2018, Bera2020}. The question of why there is a limited population of FR I XRGs remains a subject of ongoing debate \citep[see, e.g.,][]{Dennett-Thorpe2002, Saripalli2018}. A possible explanation is that after jet activity ceases, edge-brightened FR II type morphology may have relaxed into edge-darkened lobes of the FR I type, since many edge-darkened XRGs are currently experiencing renewed nuclear activity (see, e.g., Fig.~\ref{Fig:PKS2014-55}), which is seen as inner double structures of FR II type \citep{Saripalli2008,Saripalli2009,Cotton2020}. Nonetheless, there remains an unresolved question regarding the presence of a restricted number of XRGs exhibiting FR I morphologies that have not yet manifested any inner restarted activity.

Another widely accepted observation is the absence of hotspots in the wings of XRGs. While this is a well-known characteristic, it seems to challenge the notion that both the lobes and wings originate from two bidirectional outflows from binary AGN systems \citep[dual AGN model;][]{Lal2007}. In general, the wings exhibit a broad, diffuse, and lower luminosity nature, suggesting that they are more aged structures compared to the active lobes. Consequently, the plasma particles responsible for radio emission in wings, evolving under a weak magnetic field (several $\mu$G), have significantly cooled, primarily emitting in lower frequencies than the active lobes. To gain a more profound understanding of the morphological extent of these structures, to refine models of the micro-physical processes occurring within the broad cocoon structures, and to explore the interplay between bent jetted structures and the large-scale environment, high-resolution and sensitive low-frequency observations are therefore imperative \citep[][]{Hardcastle2019,Yang2019}.

The investigation of jetted structures extends to multi-wavelength analyses to gain deeper insights into the micro-physical processes within the lobes. For instance, \citet{Kraft2005} conducted a comprehensive study employing radio (VLA), optical (HST), and X-ray (Chandra) observations on 3C 403 to model the compact components of the active lobes. Their findings pointed to a synchrotron origin, positioning it as a notable example of synchrotron X-ray emission from the jet of a potent narrow-line radio galaxy \citep[see also the case of 3C 433:][]{Miller2009}. Furthermore, their examination of the inhomogeneous diffuse X-ray emission near the western lobe and wing suggested an Inverse-Compton origin involving Cosmic Microwave Background photons, assuming a moderate departure from equipartition. Such detailed modeling is crucial, especially utilizing X-shaped sources, which exhibit both the active and passive phases of jet evolution. This helps verify various assumptions used to quantify micro-scaled physics of jetted sources, connecting the emission properties.

\begin{figure*}[h!]
\centering
\includegraphics[width=0.98\textwidth]{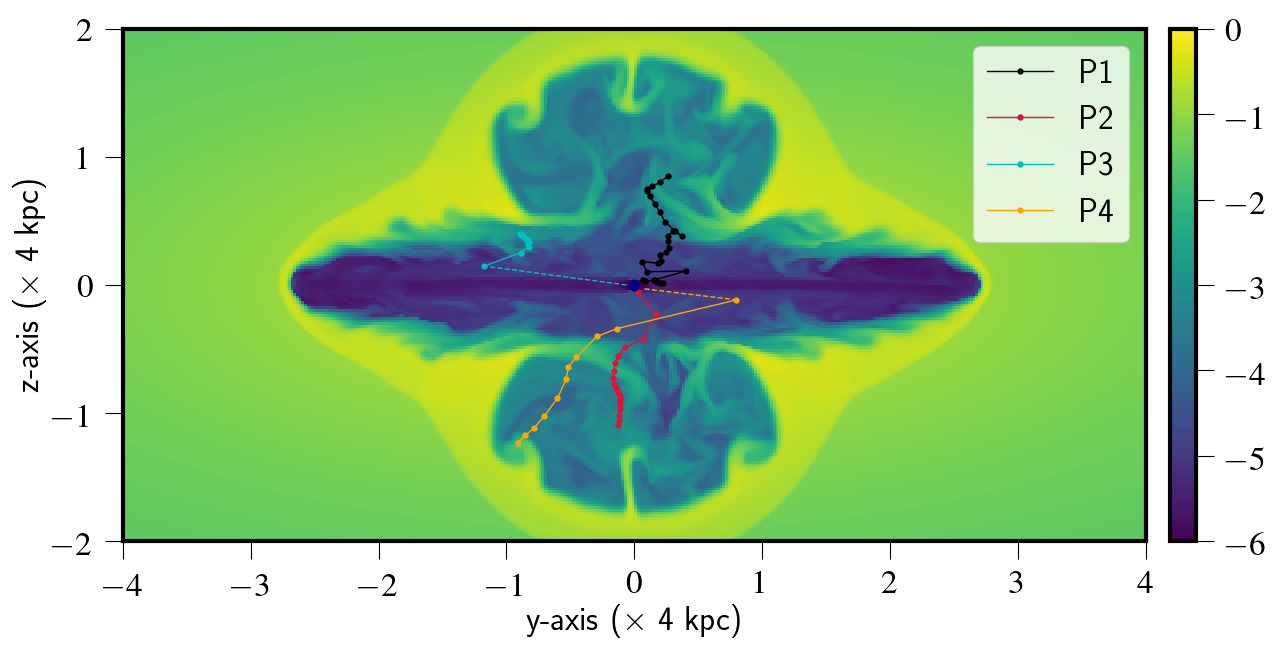}
\includegraphics[width=\textwidth,height=0.5\textwidth]{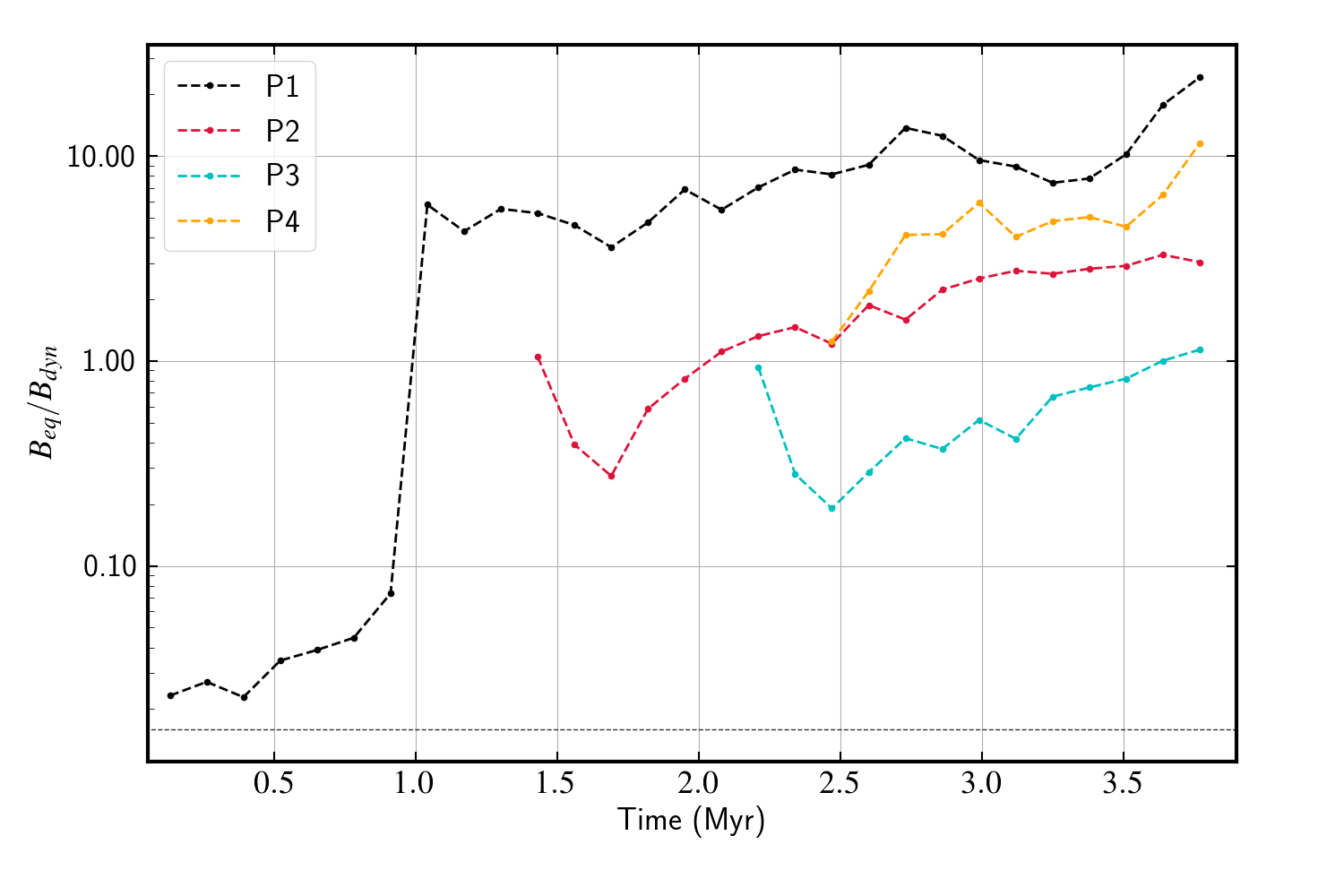}
\caption{\textit{(Top panel:)} \textit{Top:} Density (${\rm log \rho}$) slice from a 3D simulation illustrating the formation of an X-shaped structure, displaying turbulent density evolution within the cocoon and at the cocoon-ambient medium interface (at 3.78 Myr). The internal turbulence, as evidenced by the trajectories of four plasma blobs tracked since their injection, generates random shock sites where particles become energized. \textit{(Bottom:)} Evolution of the equipartition-to-dynamic magnetic field ratio ($B_{\rm eq}/B_{\rm dyn}$) for the same four plasma blobs since their injection. A value of $B_{\rm eq} = B_{\rm dyn}$ indicates true equipartition between radiating electrons and the magnetic field of the radio structure. The diverse evolution and small-scale variations observed in the graph reflect the influences of various micro-physical processes, including adiabatic cooling and diffusive shock acceleration \citep{Giri2023IAUS}. Image reused with permission; \copyright\ The authors, and The Cambridge University Press.}
\label{Fig:equipartition}
\end{figure*}

One notable assumption is the energy equipartition, which posits an equal share of energy in the magnetic field and in the particles
within the cocoon that has been built up by the jet structure \citep{Hardcastle2002}. 
This assumption is frequently employed to determine parameters such as magnetic field strength in the lobes and their radiative ages, 
subsequently governing the determination of the jet power involved and the radio luminosity. 
Note, however, that a slight deviation from the equipartition condition in a radio galaxy can lead to substantially different
spectral and dynamical ages, consequently influencing the determination of other crucial parameters \citep{Croston2005,Mahatma2020}. 
Studies such as those by \citet{Hodges-kluck2010B} have brought attention to potential discrepancies between dynamical and spectral ages 
in XRGs, a concern further addressed numerically in models proposed by \citet{Giri2022B} and \citet{Giri2022A}. 
These modeling efforts provide a more profound understanding of the micro-physical processes, including various cooling and particle 
re-acceleration mechanisms, at play within these morphological structures (see, e.g., Fig.~\ref{Fig:equipartition}). 
Such insights become particularly valuable for interpreting the anomalous spectral gradients observed in several XRGs 
(further discussed in the next section). 
Yet a deficiency remains in comprehensive modeling, in particular of high-resolution observations of XRGs, and numerical simulations on larger scales are still lacking. 

\subsection{Spectral Index Maps} \label{Sec:Spectral Index Maps}
Developing spectral index maps is crucial for gaining deeper insights into the particle evolution physics in the radio cocoon, providing key information about particle cooling and re-energization. One of the earliest works on spectral analysis of a rather unusual X-shaped source, 4C 18.68, was reported by \citet{Gower1982A}, identifying the central active-lobe like component having a flat spectrum. The spectral values tended to be much steeper in the wing and showed a tendency to steepen towards the wing edges. This was due to the effect of radiative and adiabatic cooling effects operated on the particles \citep[e.g.,][]{Kardashev1962,Longair1973,Jaffe1973} and had been noticed in a number of extended sources by then. A detailed analysis of NGC 326 by \citet{Murgia2001} further pointed out a similar trend, with wings getting even steeper with higher frequency choices \citep[an effect of particle cooling;][]{Fan2008}. A consistent observation of this spectral pattern in the wing-lobe structure has been documented in numerous studies, including recent investigations employing high-resolution and sensitive mappings of XRGs \citep{Bruno2019,Cotton2020,Mahatma2023}. Such studies have also observed a spectral gradient shifting from a flatter active lobe region to a steeper wing region, consistent with the hypothesis that the wing structure represents a relic of past AGN activity \citep[see also][]{Rubinur2017,Misra2023}.

However, we note that an earlier study by \citet{Hogbom1979} reported on XRGs 3C 192 and 3C 315, emphasizing their nearly uniform spectral distribution, calling for deeper insight into the secondary lobes. An overall radio spectral analysis (integrated over the source) of the XRG 4C +01.30, conducted by \citet{Wang2003}, further revealed a spectrum that is notably flat for an extended radio source ($0.4-5$ GHz spectral value of $\sim -0.6$). Subsequently, more sources with integrated spectral index values flatter than those of normal extended radio galaxies have been noted by \citet{Bera2020,Bhukta2022}, such as J0758 + 4406, which exhibited overall spectral values between $0.15$ and $1.4$ GHz at $-0.35$. 

The observation of such flatter spectral values of XRGs has prompted numerous studies to investigate the wing and lobe structures separately (resolving individual structures in the spectral map), revealing anomalous spectral behavior in certain sources. For example, \citet{Rottmann2001} delved into this issue by examining nine prominent XRGs, further identifying 3C 223.1 and 3C 403 as candidates where the wings exhibit flatter spectral indices compared to the active lobes and the jet hotspots. Subsequent work by \citet{Dennett-Thorpe2002} on these two sources also provided evidence of comparable or flatter spectral values in the wings than in the active lobes, with the spectral values in wing-lobe regions varying between $-0.8$ to $-0.6$.

This discovery has spurred multiple groups to conduct detailed mappings of XRGs, including the aforementioned sources, using multi-frequency observations and spectral mapping. The goal is to gain a deeper understanding of such anomalous spectral behavior, whether it is associated with particle re-acceleration mechanisms, enhanced radiation losses in the hotspots, or binary AGN scenarios capable of ejecting jets. The investigation into the origin of the anomalous spectral behavior of 3C 223.1 itself, involving different observing bands and authors, has revealed potential connections to internal micro-physical processes \citep{Gopal-Krishna2022} or the possibility of binary AGN scenarios \citep{Lal2005,Lal2007,Lal2019}.

However, analytical or numerical works to understand the possible scenario of binary AGNs ejecting jets in forming X-shaped radio structures and originating corresponding anomalous spectral behavior are yet to be conducted. A handful of numerical works have been conducted on modelling the particle evolution micro-physics in the jet cocoon of such radio galaxies. The processes of particle re-acceleration, as outlined by \citet{Fermi1949}, including diffusive shock acceleration \citep{Blandford1978,Blandford1987} and turbulent re-acceleration \citep{Rieger2007}, have been intricately modeled through numerical simulations involving powerful jets \citep{Fromm2016,Vaidya2018,Kundu2021,Mukherjee2021,Borse2021}. An investigation on the similar context, conducted by \citet{Kundu2022} has explicitly revealed the influence of particle re-energization, especially the effect of second order Fermi process, on spectral distribution. This investigation points toward the additional supply of energy to evolving particles, thereby maintaining their activity beyond what would be expected under the influence of radiative and adiabatic cooling effects (thereby showing flatter spectra for longer evolving time). Such conclusions may also be pertinent to the secondary lobes of XRGs, where particles in the wings experience re-acceleration due to internal turbulence (generates random shock sites). 

\citet{Giri2022A}, in this context, delved into these issues and demonstrated the feasibility of generating random shock sites in the wings. This occurs as particles are redirected back to the wing cavity, creating patches with flatter spectral index amidst the background of evolving cooled particles. Such phenomena significantly influence the overall wing spectra, rendering them flatter. This explanation has been referenced in elucidating the anomalous spectral behavior of 3C 223.1 \citep{Gopal-Krishna2022}. However, \citet{Giri2022A} also emphasized that the occurrence of these flatter spectra depends on the choice of frequencies. Higher frequency selections in evaluating spectral values exhibit primarily the standard behavior observed in normal radio galaxies. Substantiating this, \citet{Patra2023} recently conducted a statistical investigation to identify anomalous spectral behavior in XRGs within the frequency band of $0.14 - 1.4$ GHz. They found such anomalous behavior to be exceedingly rare, with only one sample exhibiting this peculiar trait. In contrast, the study by \citet{Lal2019}, which showcased wings comparable or flatter than the active lobes, demonstrated such behavior in a narrow frequency range of $0.24 - 0.61$ GHz, where spectral steepening is expected to be less pronounced \citep{Giri2022A,Nolting2023}. However, the exploration of a few XRGs exhibiting anomalous behavior even at higher frequency choices necessitates further study of a larger sample.

\subsection{Polarization Properties}
A notable fraction of XRGs has undergone polarization analysis to elucidate the distribution of projected magnetic field lines and fractional (linear) polarization information. Such exploration was conducted to primarily understand the dynamics of jets, internal cocoon structures and the jet-ambient medium interaction. In the active lobes, the overall behavior of magnetic field lines has been observed to align with the jet flow or the lobe edges \citep{Hogbom1979,Black1992,Johnson1995}, exhibiting a typical pattern commonly found in FR II radio sources \citep{Bridle1994}. This is consistent with a model wherein shock compression, arising from the interaction between the jet and ambient medium, plays a pivotal role in aligning the magnetic field longitudinally as shown by \citet{Liang1981} \citep[see also the recent numerical works by][]{Giri2022B,Meenakshi2023}. In this model, the jet material interacts with an ambient medium (plows into it) containing tangled fields, and the ensuing compression at the bow shock and shear at the boundary layer between the jet cocoon and the shocked ambient medium organizes the field along the lobe. The hotspots within the active lobe reveal magnetic field lines oriented perpendicular to the jet flow direction \citep{Black1992,Baghel2023}, suggesting a potential compression of field lines at the termination shock of the jet. Such a compression can be inferred by the detection of higher fractional polarization values observed in hotspots than the lobes \citep[fractional polarizartion, $\sim 40-50$ percent in comparison to $\sim 30$ percent;][]{Black1992}. Often hotspots are followed by bow shock-like sheath regions where highly ordered B-field lines are observed, displaying a distribution reminiscent of a bow-shock pattern \citep{Black1992,Johnson1995,Baghel2023}.

The distribution of magnetic field lines in the wing is observed to be even more ordered than in the active lobe, running parallel to the wing edges \citep{Hogbom1979,Black1992,Rottmann2001,Dennett-Thorpe2002,Koziel2012}. The assertion is supported by fractional polarization values in the wings, with values reaching or exceeding 50 percent. Such ordered distribution of magnetic field lines has been documented in sensitive mappings of giant XRGs, as seen in the work of \citet{Cotton2020}. 

In numerous XRGs, the smooth transition of magnetic field lines into wing-lobe structures suggests that the two pairs of lobes are likely interconnected entities, implying a shared evolutionary process rather than distinct and independent developments \citep[e.g., 3C 315, 3C 223.1, 3C 34, 3C 136.1, 3C 403, PKS 2014-55][]{Hogbom1974,Black1992,Johnson1995,Rottmann2001,Dennett-Thorpe2002,Cotton2020}. Numerical simulations in this regard have been employed to replicate the polarization patterns observed in winged sources using various models \citep{Rossi2017,Giri2022B}. Despite differences in the models, a consistent outcome has been observed i.e., the field lines within the lobes consistently align with the flow lines, posing challenges in determining the usefulness of polarization information for understanding the underlying formation mechanism. Nonetheless, a comprehensive modeling approach is crucial from both a simulation standpoint, involving the inclusion of micro-physical processes in particle evolution and large-scale XRGs simulation, and an observational standpoint, requiring high-resolution mapping to discern polarization patterns in compact components.

\section{XRG formation models and Challenges} \label{Sec:XRG formation models and Challenges}
Understanding the properties described above for X-shaped radio galaxies provides crucial insights that pave the way toward the formulation of potential formation mechanisms. The existing studies have identified two possible aspects contributing to the formation of inversion-symmetric bending. These factors are associated with either the impact of an asymmetric triaxial ambient medium through which the jet propagates or a complex mechanism occurring within the central AGN where the jet originates. 

\subsection{The Back-flow Diversion Model}
\subsubsection{The Model Setup}
Considering the association of XRGs primarily with FR type II radio galaxies, the active lobe is likely to create back-flowing plasma that flows backwards towards the center from the jet termination point. This backflow originates from a pressure imbalance at the jet head-ambient medium interface, followed by a sharp change in entropy at that point \citep{Bromberg2011, Cielo2017}. The trajectory of the back-flowing material can undergo mechanical alterations in the presence of an asymmetric, triaxial medium, directing the plasma through the influence of steepest pressure gradient force and buoyancy force \citep[e.g.,][]{Gull1973}, eventually causing an almost lateral bend. This situation is anticipated for both arms of the bidirectional jet-lobe. However, to give rise to an X-like winged structure, the back-flowing material from the lobes needs to bend in the opposite direction. \citet{Leahy1984} proposed that an ambient medium characterized by an ellipsoidal triaxial morphology could disrupt symmetry, subsequently causing the backflowing material to bend in the opposite direction. 

In a similar vein, \citet{Capetti2002} has put forth a comparable argument involving the evolution of an overpressured cocoon within an ellipsoidal triaxial medium. By following the highest pressure gradient path, the cocoon has been observed to generate an X-like morphology. This proposition gains further support from \citet{Hodges-Kluck2011} through extensive large-scale simulations (on galaxy group or cluster scales), which model the long-term evolution of XRGs, demonstrating that a higher ellipticity of the ambient environment produces a prominent XRG morphology. Subsequently, \citet{Rossi2017} undertook a more realistic modeling approach focusing on the formation phase of XRGs based on this framework (jets evolving inside a triaxial galaxy). This involved incorporating relativistic effects and magnetic fields, demonstrating the formation of various wing-lobe structures based on the propagation angle of the active jet in relation to the major axis of the ambient medium. The model has been further expanded by \citet{Giri2022A} to encompass pragmatic emission signatures, taking into account the influence of particle cooling and re-acceleration processes. 

\subsubsection{Strengths of the Back-flow Model}
It is evident that the Back-flow model relies on the lateral bend of back-flowing plasma, following the maximum pressure gradient path as influenced by the triaxial ambient medium. In the case of an ellipsoidal medium, such as a galaxy, galaxy group, or galaxy cluster with this shape, the steepest pressure gradient path aligns along the minor axis of the medium. This naturally explains why the wing structure of XRGs predominantly aligns along or exhibits a minute spread with respect to the minor axis of the ambient environment \citep{Capetti2002,Hodges-kluck2010A,Gillone2016}. In this context, the active lobe is observed to propagate along or around the major axis of the ambient medium with a spread of $\lesssim 50^{\circ}$. Such a relatively broad spread of active lobe propagation with respect to the major axis of the ambient environment results from the fact that FR-II radio galaxies, in general, show no trend in their propagation direction in relation to the host medium's axes \citep{Saripalli2009}. 

Irrespective of this, the propagating jet has been found to generate substantial backflowing plasma from the jet-head that eventually channels into the wing as it interacts with the denser ambient medium, following the path of the lowest pressure gradient (lowest rate of pressure decrement along the major axis). Notably, numerical modeling by \citet{Rossi2017} has demonstrated that an angle of $30^{\circ}$ is capable of producing the observed X-like structure based on this model. It is also anticipated that an actively propagating jet along the major axis of the ambient medium would yield prominent XRGs. Furthermore, it is consequential to understand why the wing structure lacks hotspots at their edges, and the typical extent of the wings is primarily detectable in lower frequencies \citep{Yang2019}, since they mostly consist of older cooling particles \citep{Giri2023IAUS}. The model also offers an explanation for the propensity of XRGs to originate in elliptical mediums characterized by higher ellipticity than typical FR-II sources \citep{Saripalli2009}. 

The majority of XRGs exhibiting a wing-to-active lobe length ratio less than 1 \citep{Bera2020,Bhukta2022} can be effectively elucidated by this model. This is attributable to the anticipated movement of back-flowing plasma, which is expected to travel backward from the active lobe end before inflating the wing region. This observation gains prominence from the tendency that XRGs with larger extents typically display shorter wing lengths \citep{Saripalli2009}. Notable examples illustrating such a scenario include PKS 2014-55, J0318+684, and 3C 34 \citep{Cotton2020, Bruni2021, Mahatma2023}. XRGs with wing lengths comparable to or larger than the active lobes can also be explained by the fact that the overpressured cocoon scenario can induce a supersonically moving wing \citep{Giri2023B}, which, along with the projection effect, would generate such morphology \citep[][]{Hodges-Kluck2011,Rossi2017,Giri2022A}. 

A natural explanation for the prevalence of these radio galaxies in the low-powered FR IIs can also be derived from this model. The FR I radio galaxies are believed to have minimal back-flowing plasma, as they begin to disperse and lose collimation at the jet head \citep{Massaglia2016}. A high-powered FR type II radio galaxy is expected to produce increased back-flowing plasma due to a higher pressure imbalance between the jet head and ambient medium. However, the rapid advancement of the powerful jet through the medium results in a faster increase in its length as well. Consequently, despite the substantial production of back-flowing plasma, the amount of matter channeled into the wing is reduced, resulting in a less pronounced X-shaped structure \citep{Rossi2017}. In this regard, a denser ambient environment is expected to lead to a more prominent wing as the jet propagation is hindered by increased jet-ambient medium interaction. Interestingly, X-shaped radio galaxies are not commonly observed in denser environments. This scarcity likely is a direct consequence of the sparser occurrence of FR II radio galaxies in such environments \citep[median richness of $15$;][]{Gendre2013}. Studies, such as by \citet{Joshi2019}, suggest a similar environmental richness for both FR IIs and X-shaped radio galaxies in comparison to FR Is with median richness of $\sim 30$. The residence of FR II radio galaxies in low-density environments has sparked debate, with some attributing it to central engine activity, like conditions in the accretion flow, while others link it to the radio jet's interaction during its journey through the ambient environment \citep{Lin2010,Buttiglione2010,Capetti2017}.

The model attempted to provide an explanation for winged sources exhibiting FR-I type morphology, even though they are exceedingly rare \citep{Bera2020}. Modeling by \citet{Hodges-Kluck2011} indicated that once the jet injection process decays, i.e., the AGN enters a quiescent phase, the lobes are expected to decay to luminosities more typical of FR I sources. Therefore, the limited number of identified FR I sources may indicate the phase of this special event. This argument is further supported by the reporting of \citet{Saripalli2008, Saripalli2009}, demonstrating that a notable fraction of these low-powered radio galaxies display inner doubles of FR II type, suggesting a restarting AGN activity. Furthermore, additional support for this explanation is evident from the fact that inner deviation sources, which are indicative of XRGs, are predominantly associated with FR II morphologies \citep[e.g., 36 out of 37,][]{Saripalli2018}. In contrast, outer deviation sources generating XRG candidates are found to be associated with FR Is, albeit with a lesser fraction \citep[only 4 show FR I characteristics out of 19 sources,][]{Saripalli2018}.

The recent discovery of anomalous spectral behavior in XRGs challenges the idea that wing structures are older (see Section~\ref{Sec:Spectral Index Maps}). While this questions the Back-flow model's prediction, it is crucial to consider that particle re-acceleration mechanisms can significantly re-energize flowing particles, concealing their actual age in contrast to the predictions made in spectral index maps (see Section~\ref{Sec:Spectral Index Maps} for details). Addressing how the wing structure generates such a turbulent medium to energize particles, \citet{Giri2022A} suggested that the diversion of backflowing plasma at the active lobe$-$wing base may generate turbulence, forming random shock sites in the wing where particles get re-energized. This phenomenon is reflected in the spectral index map, making the wing flatter than expected. Moreover, the modeling by \citet{Giri2022A} predicted that spectral maps with a larger frequency range would exhibit the standard behavior expected from an evolving wing, as particle cooling is extensive in higher frequency choices. The study by \citet{Patra2023} confirms this prediction, with all but one source showing this anomalous spectral behavior. While Fermi processes (both 1st and 2nd order) can be active in XRGs generated from other models, existing studies suggest that the Back-flow model is capable of explaining such a confounding property.

\subsubsection{Caveats of the Back-flow Model}
While this model may be well-suited for explaining the X-shaped morphology in inner deviation sources and in some of the potential outer deviation sources \citep[e.g.,][]{Cotton2020}, it faces challenges in elucidating the origin of prominent outer deviation sources like NGC 326 \citep{Murgia2001}, J1153.9+5848 \citep{Bruni2021}, and J1159+5820 \citep{Misra2023}. While the majority of XRGs exhibit wing alignment along the minor axis of the ambient environment, recent discoveries present instances where wings align along the major axis of the ambient medium \citep{Hodges-kluck2010B,Joshi2019}. The model also faces challenges in explaining sources with hosts of circular geometry or asymmetric distributions of stars and gas, as documented by \citet{Heckman1982, Evans1999, Saripalli2009, Misra2023}. Most importantly, the Back-flow model encounters difficulties in accounting for wings that are noticeably longer than the active lobe, as seen in sources reported like in \citet{Gower1982A, Wang2003, Bruno2019, Ignesti2020}. Despite the potential influence from projection effects, the model struggles in elucidating the collimation of such extended wings. This is because the projection effects have been noticed to broaden and diffuse the wing structure with an increase in viewing angles almost always \citep{Hodges-Kluck2011, Giri2022A}.

While the Back-flow model provides an explanation for the anomalous spectral behavior observed in several XRGs, it is yet to be firmly established as the underlying model for the majority of XRGs. This is because particle re-acceleration could also be an intrinsic component of other formation models, nonetheless, its impact on the spectral map needs to be examined.

\subsection{The Jet Reorientation Model}
\subsubsection{Proposed Reorientation Model(s)}
Another explanation for the formation of wings involves attributing the structure to a past jet reorientation event, a framework that has also been widely discussed. In the context of the jet reorientation timeline, such events can be broadly classified into two types: slow reorientation, where the jet slews gradually over several million years, and fast reorientation, where the jet flips to a new angle almost instantaneously. The motivation for such a scenario can be inferred from the studies of double-double radio galaxies, where the restarting jet has been observed to be generated along a new direction \citep[e.g.,][]{Saripalli2013,Nandi2021}. 

However, to explain the observed X-shaped morphology, the jet reorientation event has to occur at a substantial angle \citep[average alignment angle of $75^{\circ}$;][]{Capetti2002,Bhukta2022}. Analytical exercises involving a diverse set of physical processes have demonstrated that such a scenario could indeed be possible. Among these, the inbound motion and coalescence of binary SMBHs, influencing a shift in the spin direction of the jet-ejecting black hole, has gathered significant attention. This is due to its diverse range of possibilities for shedding light on topics such as the co-evolution of galaxies and SMBHs \citep{Yu2002,Bansal2017,Kharb2017}, as well as low-frequency gravitational wave astronomy \citep{Amaro-Seoane2012,Zhu2015}.

$\bullet$ {\em Precession of Binary Black Holes:} 

In the hierarchical galaxy formation model of cold dark matter cosmology, massive galaxies result from successive galaxy mergers \citep{Ji2014, Mancillas2019,Giri2023A}. During these mergers, supermassive black holes in the galactic core quickly form hard SMBH binaries at a pc-scale separation. The binaries may stall for a timescale longer than the Hubble time if the gravitational potential at the galactic nucleus is spherical and stellar relaxation is dominated by two-body scattering \citep{Begelman1980, Yu2002}. Radio observations can probe this scenario when at least one of the black holes produces a jet. The orbital motion imposes a widening helical pattern on the jet \citep[e.g.,][]{Kun2015, Britzen2017, Kharb2019, Jiang2023}. 
Supermassive black hole binary with subparsec separations can raise jet precession periods of the order of several Myr, significantly less than the typical ages of observed radio galaxies spanning several hundred Myr. This results in visible jet curvature on radio maps of larger scaled jets \citep{Krause2019}. In the context of XRGs, in-depth analyses employing jet precession models have scrutinized the curved wing-active lobe structure, seeking to delineate the parameters of potential binary black hole pairs at the center of the host galaxies \citep{Ekers1978, Gower1982B, Gong2011, Rubinur2017}. Their identification of nearly million-year jet precession has facilitated the inference of the separation, mass ratio and orbital period of potential SMBH binaries at the center. In this context, numerical simulations have been employed to model the large-scale bent structure of the jet, incorporating the evolution of jet precession. These simulations have successfully generated XRGs with a similar topology, underscoring the potential occurrence of such phenomena \citep{Giri2022B, Horton2020, Nolting2023}.

$\bullet$ {\em Coalescence of Binary Black Holes:} 

Due to the non-spherical gravitational potential induced by the rotating central object and massive perturbers from gas disks at galactic centers, the separation between binary SMBHs is further reduced, leading to their merging within a Hubble time \citep[see review by][]{Merritt2005}. 
\citet{Merritt2002} investigated the merger of binary SMBHs with a mass ratio of $M_2/M_1 = 1/4$, exploring the instantaneous change in spin magnitude $\delta S$ during the coalescence, approximated as $GM_1^2/2c$ ($G$ is gravitational constant and $c$ is speed of light). In this scenario, $\delta S$ is calculated as the difference between the resultant spin ($S$) after the coalescence of the black holes and the spin of the heavier black hole involved in the merger ($S_1$).
Introducing the parameter $\lambda = \delta S/S_1$, the study reveals that the average cosine of the realignment angle $\zeta$ can be expressed as
$\langle \cos\zeta \rangle \simeq (2/3\lambda)$ 
for $\lambda > 1$. 
Notably, if the larger black hole is initially rotating at a relatively slow rate (comparable to, for instance, $\sim 0.1\ GM_1^2/c$, thereby fulfilling $\lambda > 1$), the resulting realignment angle is approximately $\simeq 82^{\circ}$. Therefore, such a flipping of the spin axis, and consequently, the jet ejection axis, can result in the formation of an X-shaped jetted morphology. 

The recent analytical work by \citet{Garofalo2020} presents a slightly different finding in this context. It suggests the presence of low-spinning black holes in X-shaped radio galaxies. However, the change in the spin direction of the jet-ejecting black hole is initiated by the transition from a retrograde to a prograde rotating state, influenced by a recent merger.

In the above exploration of an instantaneous spin axis flip, it has been assumed that the underlying accretion disk does not fall apart, allowing the jet ejection process to persist post-reorientation. It is noteworthy that \citet{Liu2004} conducted an analytical study, emphasizing the substantial role played by the strong coupling between binary black holes and the accretion disk during black hole mergers in reorienting the spin axis of the jet-ejecting black hole. Subsequently, \citet{Liu2012}, focusing on the intricate relationship between mass ratios of colliding black holes and the spin axis flip angle, indicated that for a significant spin axis flip, the mass ratios of colliding black holes need to be comparable ($\gtrsim 0.3$), suggesting major mergers of black holes \citep[see also,][]{Gergely2009}. This implies that such a spin flip scenario may only occur in major galaxy mergers, justifying the low occurrence of XRGs given the substantial rate of galaxy mergers \citep[e.g., see][for merger rate in the local Universe]{Giri2023A}. The latter study underscore the requirement for high-spinning black holes.

$\bullet$ {\em Inhomogeneous Mass Accretion:}

The presence of matter with angular momentum, not aligned with the spin of the central SMBH, can prompt a realignment of the spin axis. 
In galaxy mergers, where the merging orbit is expected to be random, the misaligned gas and dust resulting from the merger tend to converge towards the gravitational center \citep{Ji2014,Wang2020}, influencing the stability of the accretion disk and, consequently, affecting the spin axis of the black hole. The Lense-Thirring effect, described by \citet{Lense1918}, elucidates the coupling between the spin of a rotating black hole and the angular momentum of orbiting matter.
This effect induces a torque by the central black hole on the orbiting matter, especially for orbits that are not in the equatorial plane, and feels an equal and opposite absolute torque. Therefore, the Lense-Thirring effect provides a mechanism for the alignment of the spin axis of the central SMBH with the misaligned angular momentum of the accreted matter.

\citet{Bardeen1975} applied this concept to accretion disks around rotating black holes and indicating that a portion of the misaligned accretion disk undergoes Lense-Thirring precession with an angular velocity ($\omega$) given by $\omega = 2 J/r^3$
where, $J$ represents the angular momentum of the black hole, and $r$ corresponds to the radius from the black hole. \citet{Bardeen1975} demonstrated that a critical radius exists, where the infall time of matter onto the black hole becomes equal to the orbital period. Beyond this critical radius, the precession rate $\omega$ accumulates over multiple orbits, resulting in the development of a stable, large-scale warp in the accretion disk \citep[see also,][]{Liska2019}. This warped disk, with its significant angular momentum at larger radii, eventually leads to a change in the spin axis of the supermassive black hole.

In this context, the study by \citet{Babul2013} proposes a relation connecting the maximum flip angle $\zeta_{\rm max}$ of the black hole to
the ratio between gas inflow $\Delta M_{\rm gas}$ and the black hole mass $M_{\rm BH}$, 
 \begin{equation} \label{Eq1}
 \zeta_{\rm max} \approx \tan^{-1} \left( 73.5 \frac{\Delta M_{\rm gas}}{M_{\rm BH}} \right)
 \end{equation} 
When considering a relatively modest gas inflow, such as $\Delta M_{\rm gas} \approx 3 \times 10^7 M_{\odot}$,
Equation~\ref{Eq1} shows a resulting tilt angle of approximately $\sim 65^{\circ}$ for $M_{\rm BH} = 10^9 M_{\odot}$. It is important to note that this magnitude of the tilt angle is specifically applicable to slowly rotating black holes. In the case of a rapidly spinning black hole, \citet{Liska2018} illustrated a mechanism wherein a tilted accretion disk exhibiting jet precession, when extended to large scales, can give rise to XRGs depending on the viewing direction \citep{Horton2020,Giri2022B,Nolting2023}. Further exploration, exemplified by \citet{Lalakos2022}, adds support to this hypothesis, suggesting that the jet ejection axis undergoes a substantial flip to a large angle due to inhomogeneous mass accretion, resulting in erratic wobbling.

Therefore, a large-angle flip can also occur driven by inhomogeneous mass accretion, resulting in the formation of an X-shape. This scenario is most likely associated with gas-rich minor mergers, which may not leave prominent merger signatures on the ambient galaxy of the XRG host, explaining the infrequent occurrence of merger signatures in XRG hosts. In such cases, the gas from a minor galaxy merger can be funneled to the central black hole with minimal disturbance to the host galaxy's stellar distribution \citep{Dennett-Thorpe2002}.

\subsubsection{Strengths of the Jet Reorientation Model}

The wings, which are remnants of past jetted activity, presently exist as traces of jet emissions, explaining why there are no hotspots at their edges. These observations are consistent with the jet reorientation model. The model excels in providing a natural explanation for the varying length of wings compared to active lobes, including sources with significantly longer wings, while also accounting for their collimation \citep{Giri2023B}. Additionally, it sheds light on the origins of the inner and outer deviation sources, yielding strong and candidate XRG morphologies, respectively \citep[e.g.,][]{Dennett-Thorpe2002,Gopal-krishna2003,Zier2005,Rubinur2017,Misra2023}. The discoveries of XRGs with wings aligning along the major axis of the ambient medium \citep[e.g.,][]{Hodges-kluck2010B,Joshi2019} find an explanation based on this model, as numerically shown by \citet{Giri2023B} on larger scales.

The presence of recent starburst activity (around several million years old) in the hosts of several XRGs \citep{Mezcua2012}, prominent merger signatures in the hosts of some XRGs \citep{Heckman1982,Evans1999,Hodges-kluck2010B,Misra2023}, and the identification of XRG hosts with a projected circular shape of the ambient medium \citep{Saripalli2009} provide evidence that the merger-driven jet reorientation model is a plausible explanation for the origin of XRGs. Additional support for the model may arise from the detection of S-like intrinsic shapes observed in several active lobes of XRGs, indicating a precessing jet \citep[e.g.,][]{Bruno2019,Baghel2023}, supported by seminal numerical works reproducing similar structures following the evolution of a precessing jet \citep{Nolting2023}.

\subsubsection{Caveats of the Jet Reorientation Model}

The jet reorientation model is not exempt from limitations that require further investigation. 
%One notable characteristic of XRGs, where wings tend to align along the minor axis of the ambient medium \citep{Capetti2002,Gillone2016}, cannot be accounted for through this process. Additionally, t
The reason why the host galaxies of XRGs, or the host environment in general, exhibit higher ellipticity than normal radio galaxies \citep{Saripalli2009,Hodges-kluck2010A}, lacks a comprehensive explanation from this model. In a broader context, XRGs are often found in sparser environments, where the probability of galaxy mergers decreases significantly \citep{Joshi2019}. This notion is further supported by the absence of recent starburst activity in a substantial sample of XRGs \citep{Gillone2016}, coupled with the lack of gas and dust in the nuclear region and a relatively higher temperature of the nuclear environment \citep{Landt2010}. It is essential to highlight that a dry merger involving elliptical galaxies will lack significant star formation activity and show the absence of gas and dust, nonetheless, forming the binary black hole. In this case, high-resolution observations of host galaxies are crucial to trace interaction signatures arising from such dry mergers \citep{Giri2023A}, serving as a valuable method to constrain this prediction.

The distinct characteristic of XRGs, where wings align along the minor axis of the ambient medium, finds an explanation in the binary black hole coalescence model, as indicated by \cite{Liu2004}. In a galaxy merging system, as gas is expected to settle in the galactic plane with low gravitational potential, it forms an accretion disk within that plane. According to the Bardeen–Petterson effect, plasma jets are then directed perpendicular to the accretion disc, and consequently, perpendicular to the galactic plane. Therefore, the wings of XRGs, as the remnants of past jet activity, are positioned vertically to the galactic plane (i.e., along the minor axis of the host galaxy). In this regard, the orientation of active (reoriented) jet is aligned with the rotation axis of the binary black hole and is anticipated to be randomly distributed. While it has been observed that active lobes can exhibit a spread of up to $50^{\circ}$ from the host galaxy's major axis \citep{Saripalli2009}, the wings also show a broad spread of angles ($\geq 40^{\circ}$) with respect to the major axis \citep{Gillone2016}, challenging the above idea. Furthermore, this mechanism also needs to explain the formation of wings along the major axis of the host galaxy, as observed in studies by \citet{Hodges-kluck2010B,Joshi2019} and produced in large-scale simulations by \citet{Giri2023B}. Finally, this explanation seems at odds with inhomogeneous mass accretion scenarios \citep[e.g.,][]{Babul2013,Lalakos2022}, which is also likely to generate winged sources; the distribution of gas in a galaxy merging scenario may not always align with the major axis direction \citep{Wang2020}.

It is also noteworthy that XRGs are notably absent in FR I sources and powerful radio galaxies, with their prevalence primarily observed in the low-powered FR IIs \citep{Bera2020,Saripalli2018}.  

If the jet reorientation event is the mechanism behind the formation of XRGs, it needs to be explained why such reorientation happens only once in the lifetime of an XRG. One plausible explanation is that the time elapsed since the reorientation event may influence the detectability of emission from the faded lobes, given the typical age of radio lobes is on the order of $10^7$ to $10^8$ years \citep{Herwood2017,Turner2018}. However, the challenge arises when considering the discovery of sources exhibiting three cycles of AGN restart \citep{Randall2015,Chavan2023}, making it difficult to account for the unavailability of winged sources with multiple episodes of activity. Furthermore, restarted activity has been observed to occur mostly along the preexisting lobe direction \citep{Nandi2012,Mahatma2019}, even for XRGs \citep{Saripalli2008,Saripalli2009}, indicating that a jet flip to a large angle maybe less likely event.

To illustrate the limitations of individual processes causing jet reorientation, a few situations can be considered. For example, the precessing jet producing X-like morphology has to be observed from a fortuitous line-of-sight angle \citep{Rubinur2017,Giri2022B,Horton2020,Nolting2023}. This process has then problem in explaining the observed frequency of XRGs. Although studies, such as \citet{Mezcua2012}, indicate that XRGs have higher mass black holes than normal FR II sources, results from studies like \citet{Kuzmicz2017,Joshi2019}, which includes the sources from \citet{Mezcua2012}, show evidence for lower mass black holes than FR IIs. \citet{Liu2012} also demonstrated a diverse range of black hole masses in XRG host galaxies, starting as low as ${\rm log}(M_{\rm BH}/M_{\odot})$ of 7.05. Such results may seem to contradict the binary black hole merger hypothesis. These models must also account for the observation by \citet{Saripalli2009} that XRGs with larger extents, in general, exhibit less massive wings than the active lobes \citep[e.g.,][]{Cotton2020,Bruni2021,Mahatma2023}.

While a jet re-orientation ad-hoc may nicely explain the observed XRG morphologies as detailed above, the physical processes involved in 
the flipping mechanism, and in particular in the jet launching are complicated, and its details are far from understood.
For example, jet launching is commonly attributed to spinning black holes (the Blandford-Znajek mechanism), or the collimation of disk winds
(Blandford-Payne mechanism). 
Essentially, both processes involve the existence of a strong magnetic field, threading either the disk of the black hole, removing 
energy and angular momentum from these central bodies, and re-distributing them into the outflow.
Furthermore, the launching of collimated high speed jets seem to require a certain amount of axisymmetry of the system
as magnetized disk systems without such symmetry rarely show strong jets (e.g., cataclysmic variables or pulsars).

The question arises how both the accretion disk structure as well as the magnetic field is evolving during a postulated re-orientation process.
When the central source is flipped by the processes discussed above, it will require a certain amount of time 
until the geometry of components of the central source will find back to an axisymmetric structure, and also 
until the disturbed magnetic field structure has re-established a strong poloidal flux that could again accelerate material 
to relativistic speed.
We note that while a pure hydrodynamical flow that could be in principle cut off and reconnected again, 
this is not easily possible with a magnetic field.
In ideal MHD magnetic field lines cannot be `cut'.

In resistive MHD, however, physical reconnection can happen and lead to a re-configuration of the 
magnetic field alignment.
This is a well known process, although not understood in all details and on all time scales, that can be actually 
observed in nature, for example above the solar surface.
Reconnection is also considered to be responsible for high-energy particle acceleration in the magnetospheres of
compact stars and in the coronae of accretion disks. 
It can also happen in jets and jet launching regions and may generate plasmoids ejected along the jet.
However, in all these examples mentioned, reconnection, while being an energetic process, is considered as a 
local process.

In the scenario of a MHD jet re-orientation, so the disruption and re-generation of a (resistive) MHD flow, 
reconnection must be considered on a global scale, however.
The energies involved in a global jet reconnection event are huge.
Assuming a jet radius of $\simeq 100{\rm pc}$  and a $\simeq 0.1\,{\rm mG}$ jet magnetic field
we estimate an energy release involved of $\simeq 10^{52}\,{\rm erg}$.
Simply assuming a jet disruption time scale of the order of the light crossing time\footnote{a better estimate may be the Alfv\'en time scale}
$\tau \simeq 100\,{\rm pc}/c \simeq 300\,{\rm yrs}$, we calculate a luminosity involved in such a potential catastrophic event
of about $3.5\times10^{42}\,{\rm erg\,s^{-1}}$.
This would be comparable to a quasar luminosity and could hardly be imagined to happen,
and also has not yet be observed to our knowledge.

Diffusing away the magnetic field and re-establishing a new, strong magnetic flux is a process that takes time.
Therefore, although a sudden flip may happen to the central source of the jet, 
it requires time to build up a new physical setup that allows for jet launching.
The time scale and systemic parameters will certainly depend on the source of the magnetic field.
In fact, it has been shown that an accretion disk dynamo can build up a strong, jet-launching magnetic field in a
reasonable time \citep{Stepanovs2014, Vourellis2021}, typically years for AGN, 
which would still allow for a rapid re-start of a jet flow.

\citet{Liska2018} demonstrated, with numerical simulations, that jets can be constantly emitted from a system that is hosting a miss-aligned accretion disk. 
On the other hand, these authors show that turbulent processes in the accretion disk slow down precession and alignment.
In respect of the flipping scenario, this state when precession has decayed, may correspond to the state of a new, steady jet 
ejection after the flipping process.

\section{Potential Future Prospects} \label{Sec:Potential Future Prospects}
In the following, we highlight a number of important, potential future investigations that can help to elucidate and constrict 
the viable conditions under which individual models for XRGs can be applied and can function optimally. 

$\bullet$ {\em X-ray Cavities Associated with the XRGs:}

One underexplored facet in the study of X-shaped radio galaxies is to investigate the effects of the bent jet structures on 
the surrounding environment, particularly focusing on the morphological impact in the ambient gas. 
While detailed X-ray imaging studies have explored the interaction between jets and ambient media in normal jetted sources, 
the focus on bent jets has been sparse. 
Commonly, in such studies, straight, bi-directional jets create over-pressured cocoons, displacing ambient material to form cavities surrounded by shocked shells, as evident in X-ray observations, such as 
emission depression regions enclosed by bright rims \citep{Gitti2010,Hlavacek-Larrondo2012,Hlavacek-Larrondo2015,Shin2016,Vagshette2017,Vagshette2019}. 

However, recent deep exposure maps have unveiled X-ray cavities that exhibit a notable level of complexity. 
Observations in galaxy clusters like Perseus show off-axis cavities, rims, and intricate pressure wave distributions \citep{Fabian2017}. 
Similarly complex structures are observed in other systems such as M84 \citep{Bambic2023}, NGC 5813 \citep{Randall2015}, A2052 \citep{Blanton2011}, NGC 5044 \citep{David2011}, Cygnus A \citep{Smith2002}, MS0735.6+7421 \citep{McNamara2009}. 
This complexity in cavity structures may indicate deflections in the path of the jets involved, highlighting a possible jet reorientation phenomenon. 
A limited set of numerical simulations, as well suggests that the jets in these sources have 
deviated from their anticipated straight paths through a recent jet re-slewing event \citep{Falceta-Goncalves2010, Cielo2018, Lalakos2022}. Therefore, the deviations in jet morphology not only affect the radio structure, but also leave discernible signatures on the surrounding 
medium, as they are highly intertwined.

\begin{figure*}[h!]
\centering
\includegraphics[width=\columnwidth]{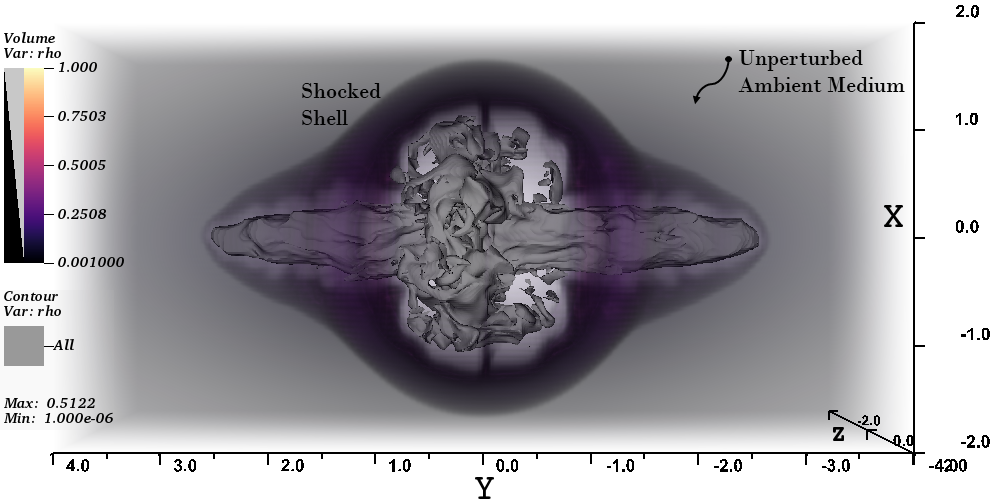}
\caption{Volume-rendered 3D representation of an XRG evolving inside a galactic environment, captured during the formation phase at 3.9 Myr. The jet material is depicted by a gray-colored contour, superimposed on the ambient material represented by the magenta color bar. This depiction illustrates the interconnected nature wherein the jet lobes push aside ambient material along their evolution path, generating cavities enclosed with shocks when observed in X-rays.  Unit density is 1 amu/cc and unit length is 4 kpc. Image reused with permission; \citet{Giri2023IAUSB}, \copyright\ The authors, and The Cambridge University Press.}
\label{Fig:density}
\end{figure*}

The intricate connection observed between jet morphology and the surrounding medium in the above scenarios may also hold true for XRGs, 
a topic that warrants detailed investigation.
In Figure \ref{Fig:density}, we depict a 3D visualization of the results of a numerical simulation capturing the initial formation phase of an X-shaped structure as it 
drills through and pushes aside the ambient gas in both the active lobe and wing directions, illustrating the interconnected nature 
of the jet lobe and ambient environment. 
The double-lobed structure of XRGs is expected to produce complex cavity signatures in the ambient medium, detection and analysis of 
which could provide evidence to constrain the underlying formation mechanism. 
This is because different levels of interaction between the wings and the ambient gas can be expected, depending on whether the structure 
forms from a back-flow model or a jet reorientation model. 

In this context, the study by \citet{Hodges-kluck2010B} illustrates the generation of such intricate structures, providing evidence for identifying multiple cavity systems associated with the XRG 4C +00.58 that has been labeled as a strong candidate 
source for jet reorientation. Later, in a low-exposure X-ray map, a prominent cavity has been associated to the wing structure in NGC 326 \citep{Hodges-kluck2012}, which has long been suggested to be another candidate with jet slewing.  
Recognizing the significance of such studies, a few numerical works have been conducted relying on jet flip model \citep{Cielo2018, Lalakos2022, Giri2023B}, further emphasizing their effectiveness in generating such structures in the ambient medium. 
The later studies, specifically focused on XRGs, have identified four clear cavity systems, as exemplified by the observational work of \citet{Bogdan2014,Ubertosi2021} in  NGC 193 and RBS 797, where the cavities align almost with a right angle to each other.

In order to discern whether such signatures differ from those produced by the double-lobed structure of XRGs in the Back-flow model, 
the study by \citet{Giri2023B} has investigated this aspect in a large-scale environment, considering the long-term evolution of the\ jetted structure for a more realistic representation of the environment of such extended sources. 
The authors highlighted that the depth and geometric alignment of the evolved cavities may serve as promising characteristics of XRGs, 
potentially contributing to the distinction of the underlying formation models. 
These intriguing findings, however, require confirmation by deep X-ray observations of the XRG ambient medium, an area that has been sparsely explored thus far.

$\bullet$ {\em Comprehensive Modelling to resolve the Caveats:}

In order to explore the formation of the X-shaped structure and its long-term time evolution, we have discussed above in detail the general behaviour anomalousities, focusing on the Back-flow and the Jet reorientation models. Highlighting the caveats of these models (see Section~\ref{Sec:XRG formation models and Challenges}), we underscore that these challenges remain open questions awaiting resolution through the application of individual mechanisms.

Tackling these issues requires an intricate numerical analysis, addressing both the formation and larger-scale evolutionary phases within a realistic environment. Additionally, comprehensive modeling of emission processes, considering the impacts of particle cooling and re-acceleration, becomes indispensable. This necessity arises because, in contrast to observational investigations, only a limited number of numerical modeling studies have been conducted, yet, providing a wealth of information on the dynamical evolution of such systems.

The absence of systematic emission modeling and long-term evolution studies for XRGs in a realistic medium remains a critical gap, but is essential for addressing the caveats discussed above. 
For instance, the study by \citet{Giri2022A} has shown the importance of incorporating micro-physical processes in addressing the anomalous spectral behavior reported in several XRGs.
Still, it is yet to be understood whether this behavior is intrinsic to all other formation models as well. 

Regarding the large-scale evolution of XRGs, hydrodynamical modeling of jet propagation in a highly ellipsoidal medium has been performed by \citet{Hodges-Kluck2011}; however, the existence of such a highly ellipsoidal environment is not always obvious.  
Recently, \citet{Giri2023B} performed a comprehensive large-scale evolution of XRGs in a less ellipsoidal triaxial medium incorporating both back-flow and jet reorientation models, highlighting detailed dynamical configurations. 
However, there is a gap in studies addressing the evolution of XRGs in a more asymmetric or nearly spherical ambient medium \citep{Black1992,Kraft2005,Saripalli2009,Hodges-kluck2012}. This aspect may pose challenges in generating XRGs from the back-flow model.
Simulation works delving into jet precession modeling have been conducted by \citet{Horton2020}, however, their pragmatic emission imprints need to be modeled for a comprehensive understanding of the evolution processes of particles inside the cocoon \citep[e.g.,][]{Nolting2023}.
This remains essential in order to explain whether jet precession is capable of addressing the observed population of XRGs along with the existence of anomalous XRGs.

In-depth observational efforts, featuring both high resolution and sensitivity, are also imperative to detect, analyze, and document additional complexities associated with X-shaped structures. 
Addressing such intricacies \citep[e.g. the presence of long-tail, and filamentary structures associated with XRGs,][see also Fig.~\ref{Fig:PKS2014-55}]{Hardcastle2019,Ignesti2020} 
demand comprehensive and complimentary observational and numerical approaches to uncover subtle features, and thereby providing valuable insights into the underlying processes shaping XRGs. 

\section{Summary} \label{Sec:Summary}
X-shaped radio galaxies represent a subset of winged radio galaxies, where two double-lobed radio structures are observed to orient at a large angle to each other, forming an inversion-symmetric configuration. 
The formation mechanism of these peculiar sources remains debatable, 
in spite of recent high-resolution and sensitive observations 
that include the analysis of both micro- and macro-scale properties, 
thus providing a diverse range of information compared to earlier studies. 

The analysis of these multi-frequency studies of such radio galaxies in fact allows for a variety of formation mechanisms, 
each of them with its own strengths and limitations, thus requiring further analytical and numerical modeling as well as more observations. 
Nonetheless, the proposed ideas can be broadly categorized into two primary hypotheses: 
one connecting the formation mechanism with a triaxial ambient medium,
and the other involving a complex activity of the central AGN that is mechanically changing the direction of the jet 
ejection. The range of properties observed in a substantial sample of XRGs until now is diverse. They may, both, support or contradict the proposed scenarios, and have thus lead to the notion that there may not be 
a universal model explaining all the properties of these radio galaxies. 
Instead, it has been suggested that different mechanisms may be at play in the individual cases.

In this comprehensive review, we have delved into these diverse characteristics of X-shaped radio structures, spanning from their large-scale ambient dynamics to the micro-physical processes governing radio cocoon evolution. 
Our exploration covered both the salient features and the confounding properties of these intriguing sources. 
Despite the wealth of observational findings, a notable gap exists in numerical modeling, particularly focusing on individual mechanisms, 
which is crucial for constraining the parameter space governing their genesis. 
This modeling is not only essential for a deeper understanding of these bent-jetted sources but also holds the key to unraveling their potential role in probing critical topics like the galaxy-SMBH co-evolution.

Our review also underscores significant deficiencies in the observational domain. The lack of detailed mapping of the larger-scale dynamics of the ambient environment where the jet terminates
hampers our ability to comprehensively study enhanced feedback mechanisms.
Additionally, the dearth of high-resolution mapping and spectral analysis of the nuclear region impedes the identification of signatures indicative of the complex AGN mechanisms at play. 
Bridging these gaps is imperative for advancing our understanding of X-shaped radio structures and determining whether they represent a distinct class of sources resulting from specific events or are a natural consequence of certain  conditions affecting normal radio galaxies.

\section*{Conflict of Interest Statement}
The authors declares no conflict of interest.

\section*{Author Contributions}
Every listed author has significantly contributed to the research, providing direct and intellectual input, and has approved the work for publication.

\section*{Funding}
G.G. is a postdoctoral fellow under the sponsorship of the South African Radio Astronomy Observatory (SARAO). The financial assistance of the South African Radio
Astronomy Observatory towards this research is hereby acknowledged (\url{www.sarao.ac.za}). 
G.B. and P.R. were supported by a INAF Theory Grant 2022, {\it Multi scale simulations of relativistic jets}, and  by the EU - Next Generation EU through the PRIN 2022 (2022C9TNNX) project. We would like to express our gratitude for the financial support extended by the Max Planck Society, facilitating a seamless publication of this manuscript.

\section*{Acknowledgement}

The MeerKAT telescope is operated by the South African Radio Astronomy Observatory, which is a facility of the National Research Foundation, an agency of the Department of Science and Innovation.

We acknowledge the use of the ilifu cloud computing facility - \url{www.ilifu.ac.za}, a partnership between the University of Cape Town, the University of the Western Cape, Stellenbosch University, Sol Plaatje University, the Cape Peninsula University of Technology and the South African Radio Astronomy Observatory. The ilifu facility is supported by contributions from the Inter-University Institute for Data Intensive Astronomy (IDIA - a partnership between the University of Cape Town, the University of Pretoria and the University of the Western Cape), the Computational Biology division at UCT and the Data Intensive Research Initiative of South Africa (DIRISA).
\bibliographystyle{Frontiers-Harvard} 
\bibliography{test}

\end{document}